# Pressured-induced superconductivity extending across the topological phase transition in thallium-based topological materials TlBi(S$_{1-x}$Se$_x$)$_2$

Cuiying Pei[1,12], Peihao Huang[2,3,12], Peng Zhu[4,5,6,12], Linlin Liu[2,3,12], Qi Wang[1,7], Yi Zhao[1], Lingling Gao[1], Changhua Li[1], Weizheng Cao[1], Jian Lv[2,3], Xiang Li[4,5,6], Zhiwei Wang[4,5,6*], Yugui Yao[4,5], Binghai Yan[8*], Claudia Felser[9], Yulin Chen[1,7,10], Hanyu Liu[2,3*], Yanpeng Qi[1,7,11,13*]

[1] School of Physical Science and Technology, ShanghaiTech University, Shanghai 201210, China
[2] State Key Laboratory of Superhard Materials and International Center for Computational Method and Software, College of Physics, Jilin University, Changchun 130012, China
[3] International Center of Future Science, Jilin University, Changchun 130012, China
[4] Centre for Quantum Physics, Key Laboratory of Advanced Optoelectronic Quantum Architecture and Measurement (MOE), School of Physics, Beijing Institute of Technology, Beijing 100081, China
[5] Beijing Key Lab of Nanophotonics and Ultrafine Optoelectronic Systems, Beijing Institute of Technology, Beijing 100081, China
[6] Material Science Center, Yangtze Delta Region Academy of Beijing Institute of Technology, Jiaxing, 314011, P. R. China
[7] ShanghaiTech Laboratory for Topological Physics, ShanghaiTech University, Shanghai 201210, China
[8] Department of Condensed Matter Physics, Weizmann Institute of Science, Rehovot 7610001, Israel
[9] Max Planck Institute for Chemical Physics of Solids, Dresden 01187, Germany
[10] Department of Physics, Clarendon Laboratory, University of Oxford, Parks Road, Oxford OX1 3PU, UK
[11] Shanghai Key Laboratory of High-resolution Electron Microscopy, ShanghaiTech University, Shanghai 201210, China
[12] These authors contributed equally
* Corresponding author's email address: Y.Q. (qiyp@shanghaitech.edu.cn) or H.L. (lhy@calypso.cn) or B.Y. (binghai.yan@weizmann.ac.il) or Z.W. (zhiweiwang@bit.edu.cn)

**SUMMARY**

**The coexistence of superconductivity and topology holds the potential to realize exotic quantum states of matter. Here we report that superconductivity induced by high pressure in three thallium-based materials, covering the phase transition from a normal insulator (TlBiS$_2$) to a topological insulator (TlBiSe$_2$) through a Dirac semimetal (TlBiSeS). By increasing the pressure up to 60 GPa, we observe superconductivity phase diagrams with maximal $T_c$ values at 6.0~8.1 K. Our density-functional theory calculations reveal topological surface states in superconductivity phases for all three compounds. Our study paves the path to explore topological superconductivity and topological phase transitions.**

---

[13] Lead Contact's email address: Yanpeng Qi (qiyp@shanghaitech.edu.cn)

## INTRODUCTION

The discovery of topological states and topological phase transitions has reshaped our understanding of physics and materials over the past few years.[1; 15; 28; 40; 41; 56; 58] The topological transition from a normal insulator to a topological insulator (TI) can be realized by chemical or mechanical pressure. The topological transition was first predicted[59] and demonstrated[44; 57] in Tl-based chalcogenides TlBi($S_{1-x}Se_x$)$_2$ by increasing the Se component. The chemical pressure increases the amplitude of the spin-orbit coupling (SOC) with the critical transition point at x = 0.5, that is, TlBiSSe, which is a 3D Dirac semimetal.[5; 23; 24; 30; 38; 43; 46] Instead, mechanical pressure pushes the bandgap of a normal insulator to close and re-open, leading to a nontrivial topological state. Furthermore, superconductivity may emerge with pressure and interplay with topological surface states, leading to possible topological superconductivity on the surface by the self-proximity effect.[8]

In this study, we apply hydrostatic pressure to TlBi($S_{1-x}Se_x$)$_2$ and investigate the effects of chemical and mechanical pressures on the phase transition and induced superconductivity, as shown in **Figure 1**. Maximum critical temperatures, $T_c$, of 8.1, 7.0 and 6.0 K are observed for TlBiS$_2$, TlBiSeS and TlBiSe$_2$, respectively. Our theoretical calculations demonstrate the coexistence of topological features and superconductivity in TlBi($S_{1-x}Se_x$)$_2$ (x = 0, 0.5 and 1) upon compression.

## RESULTS AND DISCUSSION

### Crystal structure of TlBi($S_{1-x}Se_x$)$_2$ (x = 0, 0.5, 1) in ambient condition

The TlBi($S_{1-x}Se_x$)$_2$ (x = 0, 0.5, 1) single crystals used in this study were grown using the melting method. All samples form a rhombohedral crystal structure with the space group $R\bar{3}m$, which possesses real-space-inversion symmetry and can be viewed as a distorted NaCl structure with four atoms in the primitive unit cell. The stacking sequence of each layer is -Tl-X-Bi-X- along the [111] direction, and the binding between the layers is rather strong, in contrast to the van der Waals type coupling of tetradymite semiconductors. The high crystallinity and homogeneity of our single crystals were confirmed by X-ray diffraction (XRD) and energy dispersive X-ray spectroscopy (EDS) analysis (Figure S1).

**Structural phase transition of TlBi(S$_{1-x}$Se$_x$)$_2$ (x = 0, 0.5, 1) under high pressure**

Before the physical property measurements, we investigated the pressure-induced structural evolution of TlBi(S$_{1-x}$Se$_x$)$_2$ (x = 0, 0.5, 1) using XRD with a wavelength of $\lambda$ = 0.6199 Å. Figure 2a shows a typical XRD pattern of TlBiSeS up to 40.6 GPa. Two structural phase transitions were observed under high pressure. All the diffraction peaks at 1 atm can be indexed well to a rhombohedral $R\bar{3}m$ structure by Rietveld refinement with lattice parameters of $a$ = 4.205(2) Å and $c$ = 21.734(1) Å (Figure S2a and Table S1). There was no obvious spectral change below 8.6 GPa. At 8.6 GPa, additional diffraction peaks marked by asterisks emerged, signifying the formation of phase II. When the pressure reaches 19.3 GPa, phase II began to transition into a new phase III. Above 26.6 GPa, the transformation is complete and no further transitions are observed up to 40.6 GPa. Pressure-dependent Raman spectroscopy of TlBiSeS (Figure 2c) further demonstrated the crystallographic structural phase transition sequence under high pressure, which is consistent with *in-situ* XRD results.

To identify the crystal structures of TlBiSeS under high pressure, we performed an additional structure prediction for this system using our developed structure prediction methodology.[10; 50; 51] As a result of extensive simulations, our structure searches successfully reproduced the experimental $R\bar{3}m$ structure (Phase I) at ambient pressure, validating our computational scheme. Our computed enthalpy difference curves for the predicted phases, as shown in Figure S2, indicate a phase transition at 9.1 GPa, where Phase II (space group $C2/m$ structure, No. 12) demonstrates a lower enthalpy than that of phase I ($R\bar{3}m$ structure). As the pressure is higher than 21.3 GPa, another newly predicted phase (Phase III, space group $P4mm$, No. 99) is energetic and stable. The experimental structural phase transition coincided with the theoretical prediction. In addition, the phonon dispersion curves were used to study the dynamic structural stability of the predicted structure. As shown in Figure S3, no imaginary frequency was found for these two structures, indicating their dynamic stability.

The observed XRD data were then refined using the predicted structures. Figure S4b and S4c show the typical Rietveld refinement results for TlBiSeS at various pressures. The excellent Rietveld fitting of the predicted structures makes the determination of the high-pressure phases unambiguous. Phase II possesses a monoclinic structure, whereas phase III crystals are in a tetragonal phase. Figure 2b shows the volume data as a function of pressure for the different phases. The two-phase transitions are characterized by first orders accompanying 2.4 and 4.4% volume collapse at the transitions. A similar structural evolution under pressure was observed for TlBiS$_2$ and TlBiSe$_2$ (Figure S5).

To gain insight into the structural evolution, pressure-induced coordination of Tl and Bi was derived from the Rietveld refinements, as shown in Figure 2d. In ambient condition, TlBiSeS exhibits inversion symmetry, where both Tl and Bi act as inversion centers.[60] Continuous bonding was found perpendicular to the layers, and octahedral TlSe(S)$_6$ and BiSe(S)$_6$ coupling interlayers led to an essentially three-dimensional crystal structure.[4] Although the stacking order is similar between phase I and phase II, there are seven Se(S) atoms as ligands in phase II, in contrast to the six-coordinated phase I. In phase III, Tl and Bi atoms are centered in the cuboid ($a = b \neq c$) to form eight-fold bonding with Se(S). The TlSe(S)$_8$ cuboid and BiSe(S)$_8$ cuboid are face shared and sandwiched along the $c$-axis.

**Pressure-induced superconductivity in TlBi(S$_{1-x}$Se$_x$)$_2$ (x = 0, 0.5, 1)**

At ambient pressure, TlBiSeS exhibited metallic-like behavior and electron-type charge carriers with a concentration of $n_e \sim 1.94 \times 10^{18}$ cm$^{-3}$ at 300 K (Figure S6), which is close to previously reported values.[31] Since TlBiSeS exhibits multi-phase transitions under high pressure, a question arises naturally: what will happen for transport properties? Hence, we measured the electrical resistivity $\rho(T)$ of the TlBiSeS single crystals at various pressures. Figure 3 shows the typical $\rho(T)$ curves of TlBiSeS for pressures up to 54.6 GPa. The resistivity of TlBiSeS first increases with applied pressure and reaches the maximum value at 6.3 GPa. TlBiSeS shows insulator or semiconducting-like behavior at this pressure region. Upon further increasing the pressure, resistivity starts to decrease rapidly, and a metal-insulator transition occurs. Superconductivity appears with pressure increases, and a maximum $T_c$ of 6.7 K is attained at $P$ = 19.9 GPa (Figure 3). Beyond this pressure, $T_c$ decreases slowly, but superconductivity persists up to the highest experimental pressure of ~ 55 GPa. The appearance of superconductivity in TlBiSeS is further corroborated by the resistivity data in applied magnetic fields. As shown in Figure 3e, the superconducting transition gradually shifts toward lower temperatures with increasing magnetic fields. The value of $\mu_0H_{c2}(T)$ was estimated to be 2.86 T at 23.3 GPa (Figure 3f), which yields a Ginzburg–Landau coherence length $\xi_{GL}(0)$ of 10.74 nm. In analogy with that of TlBiSeS, pressure-induced superconductivity with a maximum $T_c$ of 8.1 K and 6.0 K is observed in $TlBiS_2$ and $TlBiSe_2$, respectively. The values of $T_c$ shown here are quite higher than that in $TlBiTe_2$.[16] Temperature dependence of the upper critical field for $TlBiS_2$ at 22.7 GPa and $TlBiSe_2$ at 16.3 GPa are also shown in Figure 3f.

**T-P phase diagram of $TlBi(S_{1-x}Se_x)_2$ (x = 0, 0.5, 1)**

High-pressure transport measurements on $TlBi(S_{1-x}Se_x)_2$ (x = 0, 0.5, 1) single crystals for several independent runs provided consistent and reproducible results (Figures S7-S16), confirming the intrinsic superconductivity under pressure. Based on the above resistivity, XRD, and Raman measurements, the *T-P* phase diagram is summarized in Figure 4. These results demonstrate that high pressure dramatically alters both crystal and electronic structures of $TlBiS_2$, TlBiSeS, and $TlBiSe_2$. In phase I, the resistivity of $TlBiS_2$ exhibited a non-monotonic evolution with increasing pressure. Over the entire temperature range, the resistivity is first suppressed with the applied pressure and reaches a minimum value at approximately 4.1 GPa. With a further increase in pressure, the resistivity increases again, and a transition from metallic to semiconducting behavior occurs. This peculiar behavior of resistivity is not associated with a structural phase transition because high-pressure XRD studies revealed structural stability in this pressure range. Our high-pressure Raman results indicated a clear anomaly in the phonon linewidths of $E_g$ mode at pressures ~2.8 GPa, which is consistent with the pressure of the minimum resistivity (Figure S17). This anomaly is evidence of topological phase transitions. A similar phenomenon was observed for black phosphorus[12], $Sb_2Te_3$[11], and BiTeI[39]. A dome-shaped evolution of electrical resistivity with pressure was observed in TlBiSeS; however, the resistivity of $TlBiSe_2$ exhibited a monotonic decrease with increasing pressure. In phase II, our band structure calculations demonstrate that both $TlBiS_2$ and TlBiSeS are normal semiconductors with narrow bandgaps ($\Delta E \sim$ 238 meV and 85 meV for $TlBiS_2$ and TlBiSeS, respectively), whereas $TlBiSe_2$ is a semimetal without a global bandgap, in agreement with our resistivity results (Figure S18). The resistivity abruptly decreased and ultimately underwent metallization in this pressure range. It should be noted that a small drop of $\rho$ is observed, and no zero resistivity is achieved in this pressure region, probably because of the small amount of phase III. In phase III, the resistivity of all three samples did not change significantly in response to further increases in the pressure. Superconductivity was observed in this region, and the $T_c$ values of all three samples decreased monotonically with increasing pressure. The observed superconductivity in the high-pressure range is associated with structural instability and/or pressure-induced structural phase transition. The evolution of superconductivity is probably due to a reduction of density of state at the Fermi level.[19; 32]

**Pressure-induced topological properties evolution on $TlBi(S_{1-x}Se_x)_2$ (x = 0, 0.5, 1)**

Because superconductivity was observed in Phase III, we performed detailed *ab initio* calculations to identify the topological nature of TlBiSeS at 30 GPa. As shown in Figure 5, there is an SOC-induced band inversion near the Fermi surface at the $\Gamma$ and $X$ points. The inverted bandgaps were approximately 688 and 662 meV, respectively. We confirmed the non-trivial topological character by calculating the $Z2$ topological invariant, parity eigenvalues (Table S2). Wannier charge center (Figure S19) and topological edge states (Figure 5c) of TlBiSeS at 30 GPa[7; 9]. The inherent metallicity of TlBiSeS at 30 GPa can be easily identified by their bands crossing at the Fermi level, which are consistent with their Fermi surfaces shown in Figure 5d. Thus, TlBiSeS at 30 GPa can be interpreted as a non-trivial topological metal. Superconducting transitions in this phase are supported by their intrinsic metallic features. The Dirac-cone-like surface states of TlBiSeS at 30 GPa are located at 0.19 eV above Fermi level, inside the SOC gap between the bulk valence band and the bulk conduction band. We also preformed theoretical calculations for $TlBiS_2$ and $TlBiSe_2$, and all samples showed nontrivial band structures in phase III (Figures S20-S22). Therefore, the combined theoretical calculations and *in-situ* high pressure measurements demonstrate the coexistence of topological feature and superconductivity in $TlBi(S_{1-x}Se_x)_2$ (x = 0, 0.5, and 1) upon compression. Our study will stimulate further studies, such as the 4π-periodic Josephson effect[25; 54] and quantum oscillations under high pressure[2; 6], to investigate topological superconductivity and help the exploration of Majorana fermions.

We discovered pressure-induced superconductivity in thallium-based III-V-IV$_2$ ternary chalcogenides $TlBi(S_{1-x}Se_x)_2$ (x = 0, 0.5, 1) topological materials by combining experimental and theoretical investigations. High pressure dramatically alters the electronic state, and superconductivity is observed in all three samples after the two structural phase transitions. Theoretical calculations indicated that $TlBi(S_{1-x}Se_x)_2$ (x = 0, 0.5, 1) with a tetragonal *P4mm* structure is considered as a topological metal, which coexists with superconductivity. Our results demonstrate that $TlBi(S_{1-x}Se_x)_2$ compounds with a nontrivial topology of electronic states display new ground states upon compression and have potential applications in next-generation spintronic devices.

**EXPERIMENTAL PROCEDURES**

*Resource availiability*

*Lead Contact*

Further information and requests for the samples should be directed to and will be fulfilled by the Lead Contact, Yanpeng Qi (qiyp@shanghaitech.edu.cn).

*Materials Availability*

All the samples used in this study are available from the Lead Contact without restriction.

*Data and Code Availability*

All relevant data are available from the corresponding author upon reasonable request.

Sample Preparation

Single crystals of TlBi($S_{1-x}Se_x$)$_2$ (x = 0, 0.5, 1) were grown by a melting method starting from high purity (more than 4N) of elementary Tl, Bi, Se and S shots sealed in an evacuated quartz tube. In order to obtain crystals with high-quality, we performed surface cleaning for all starting materials except S, to remove the oxide layers formed in air.[52; 53] The elements were heated to 800 °C in several hours and kept for 48 h, at which the tubes were intermittently shaken to ensure homogeneity of the melt; after that, they were slowly cooled down to 400 °C in 100 h and subsequently down to room temperature with the furnace switched off.

*High-pressure Measurements*

Nonmagnetic diamond anvil cell (DAC) was employed to perform the *in situ* high-pressure resistivity measurements as described elsewhere.[3; 34-36; 42] A cubic BN/epoxy mixture was used as insulating layer between BeCu gaskets and electrical leads. Four Pt foils were arranged in a van der Pauw four-probe configuration to contact the sample in the chamber for resistivity measurements. Pressure was determined by the ruby luminescence method.[27] The *in situ* high-pressure Raman spectroscopy measurements were performed using a Raman spectrometer (Renishaw inVia, U.K.) with a laser excitation wavelength of 532 nm and low-wavenumber filter. Symmetric DAC with anvil culet sizes of 300 μm was used, with silicon oil as pressure transmitting medium (PTM). *In situ* high-pressure synchrotron x-ray diffraction (XRD) measurements were carried out at room temperature with sample powder grinded from single crystals at the beamline BL15U of Shanghai Synchrotron Radiation Facility (X-ray wavelength λ = 0.6199 Å). Symmetric DACs with anvil culet sizes of 300 μm and T301 gaskets were used. Silicon oil was used as the PTM and pressure was determined by the ruby luminescence method.[27] The 2D diffraction patterns were analyzed using the FIT2D software.[14] General Structure Analysis System (GSAS) and the graphical user interface EXPGUI[26; 47] were employed for Rietveld refinements on crystal structures under high pressure.

*Theoretical Calculation*

A swam-intelligence-based CALYPSO method and its same-name code[10; 50; 51] was used to predict structure. VASP plane-wave code[20; 21] was employed to calculate Density functional total energy and structure relaxation. Perdew-Burke-Ernzerhof (PBE) generalized gradient approximation density functional[37] and frozen-core all-electron projector-augmented wave (PAW) potentials[22] was adopted in our calculations. The electronic wave functions are expanded in a plane-wave basis set with a kinetic energy cutoff of 400 eV. The Brillouin zone (BZ) are sampled with a k-meshes with a reciprocal space resolution of $2\pi \times 0.03$ Å$^{-1}$ and energies are converged to 1 meV/atom. Supercell approach[33] as implemented in the Phonopy code[48] were used to calculate phonon spectrum. Heyd-Scuseria-Ernzerhof (HSE06) Hybrid functional implemented in PWmat[17; 18] to improve the accuracy of the band gap. It runs on graphics processing unit (GPU) processors. The NCPP-SG15-PBE[13; 45] pseudopotential is used as well. Using an energy cutoff of 50 Ry, the convergence threshold of the total energy is less than $10^{-4}$ eV. The Brillouin zone sampling is performed on Monkhorst–Pack scheme on 4*6*9 special k-points for structure optimization and electronic structural calculations. The maximally localized Wannier functions were constructed from DFT by the WANNIER90[29] code and then used to calculate the Wannier charge center, the Z2 topological invariant, and edge states. Based on tight-binding parameters from Wannier90[29], WannierTools[55] provides post-processing analysis. Band topologies were calculated using vasp2trace code[49].

**SUPPLEMENTAL INFORMATION**

Supplemental information can be found online.

**ACKNOWLEDGMENTS**

C. Pei, P. Huang, P. Zhu and L. Liu contributed equally to this work. We thank Prof. Yanming Ma for valuable discussions. This work was supported by the National Key

R&D Program of China (Grant No. 2018YFA0704300, 2020YFA0308800), the National Natural Science Foundation of China (Grant No. U1932217, 11974246, 12004252, 92065109 and 12074138), the Science and Technology Commission of Shanghai Municipality (19JC1413900) and Shanghai Science and Technology Plan (Grant No. 21DZ2260400). The authors thank the support from Analytical Instrumentation Center (SPST-AIC10112914) and Centre for High resolution Electron Microscopy (C$h$EM) (No. EM02161943), SPST, ShanghaiTech University. The authors thank the staffs from BL15U1 at Shanghai Synchrotron Radiation Facility for assistance during data collection. B.Y. acknowledges the financial support by the European Research Council (ERC Consolidator Grant "NonlinearTopo", No. 815869) and the ISF - Quantum Science and Technology (No. 1251/19). Z.W. thanks the Analysis & Testing Center at BIT for assistance in facility support.

## AUTHOR CONTRIBUTIONS

Y.P.Q. designed research; C.Y.P., P.Z., Q.W., Y.Z., L.L.G., C.H.L., W.Z.C., X.L. performed experiments; P.H.H., L.L.L., J.L. carried out theoretical calculations; Y.P.Q., H.Y.L., B.H.Y., Z.W.W., C.Y.P., P.H.H, L.L.L., Y.G.Y., C.F. and Y.L.C. analyzed data and wrote the paper.

## DECLARATION OF INTERESTS

The authors declare no competing interests.

**FIGURE TITLES**

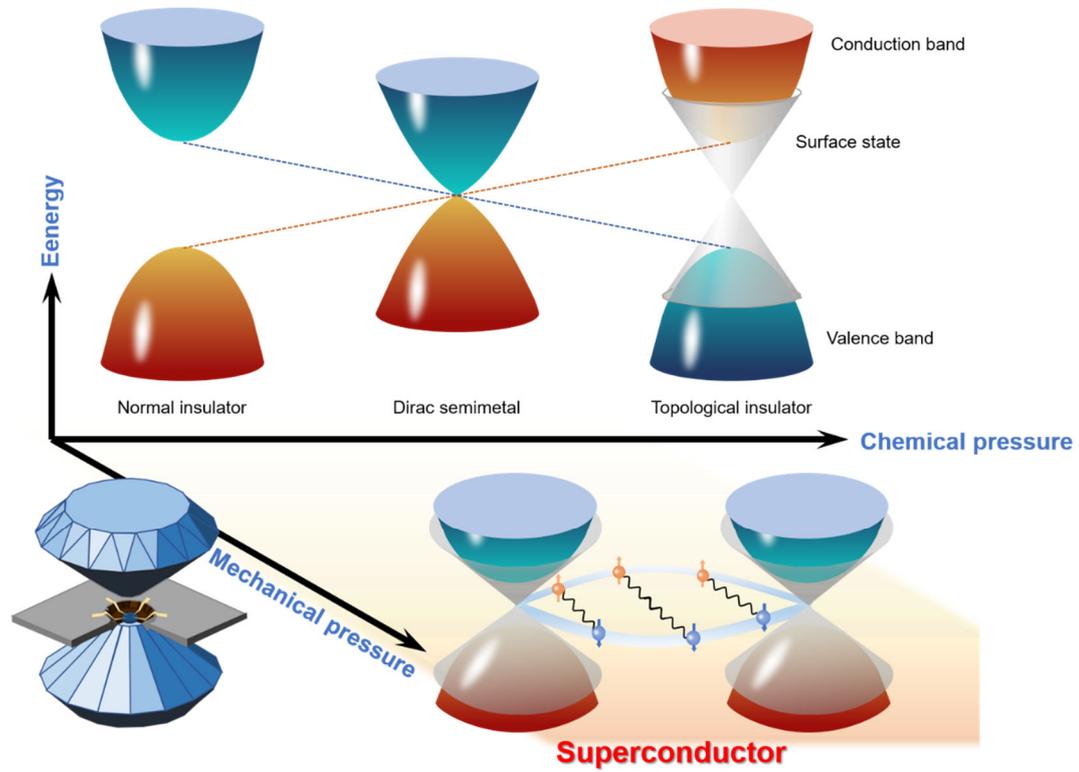

**Figure 1.** Schematic of chemical pressure and mechanical pressure effect on TlBi(S$_{1-x}$Se$_x$)$_2$ compounds.

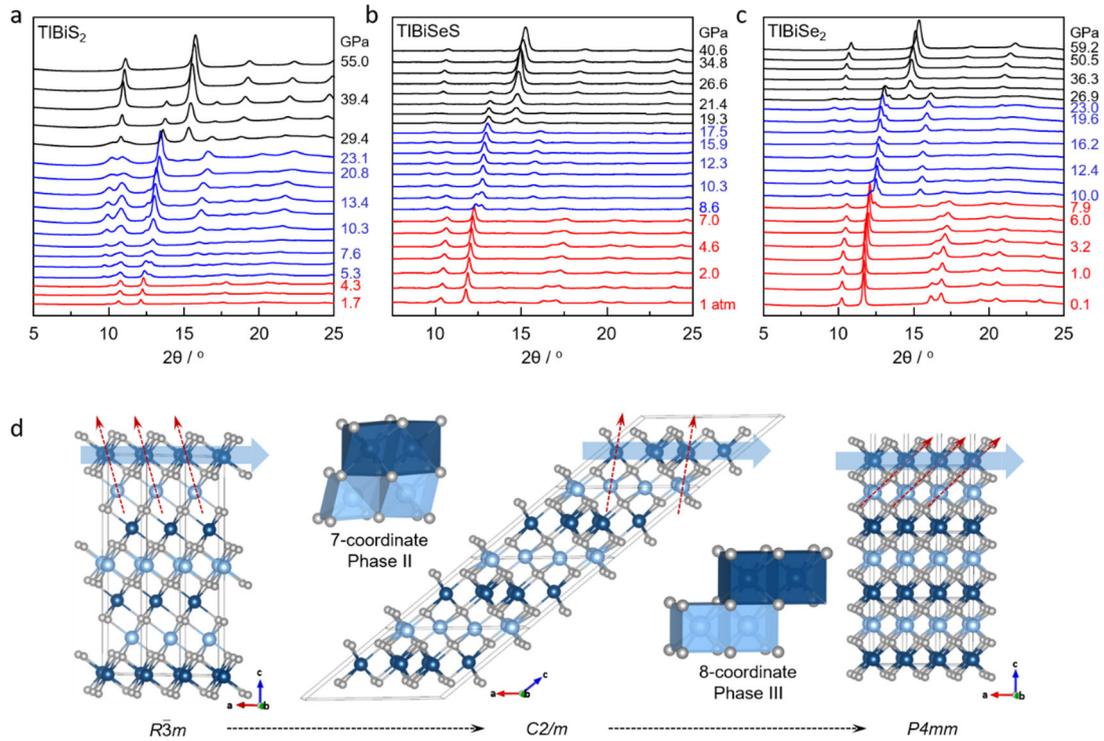

**Figure 2.** Crystal structure evolution under high pressure. (a) XRD patterns of TlBiSeS under pressure at room temperature with an x-ray wavelength of $\lambda = 0.6199$ Å. The red, blue, and black patterns distinguish phase transition from phase I ($R\bar{3}m$) to phase II ($C2/m$) and phase III ($P4mm$), respectively; (b) Pressure-dependence of volume for TlBiSeS in phase I ($R\bar{3}m$), phase II ($C2/m$), and phase III ($P4mm$); (c) Raman spectra of TlBiSeS under pressure at room temperature; (d) Crystallographic structure of the three phases of TlBiSeS under various pressures.

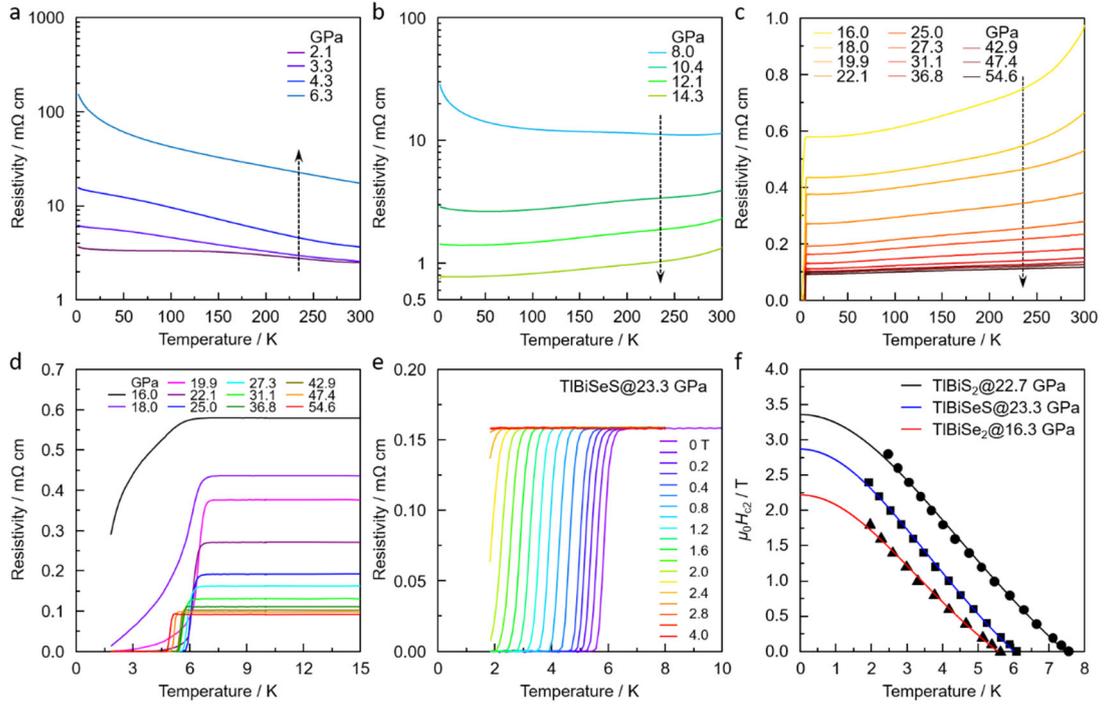

**Figure 3.** Evolution of the electrical resistivity as a function of pressure. Electrical resistivity of TlBiSeS as a function of temperature for various pressures in run II (a, b, and c); (d) Temperature-dependent resistivity of TlBiSeS in the vicinity of the superconducting transition in run II; (e) Temperature dependence of resistivity under different magnetic fields for TlBiSeS at 23.3 GPa in run III; (f) Temperature dependence of upper critical field for TlBiS$_2$ at 22.7 GPa, TlBiSeS at 23.3 GPa and TlBiSe$_2$ at 16.3 GPa, respectively. $T_c$ is determined as the 90% drop of the normal state resistivity. The solid lines represent the Ginzburg–Landau (G-L) fitting. The $\mu_0 H_{c2}(0)$ is 3.36, 2.86, and 2.22 T, respectively.

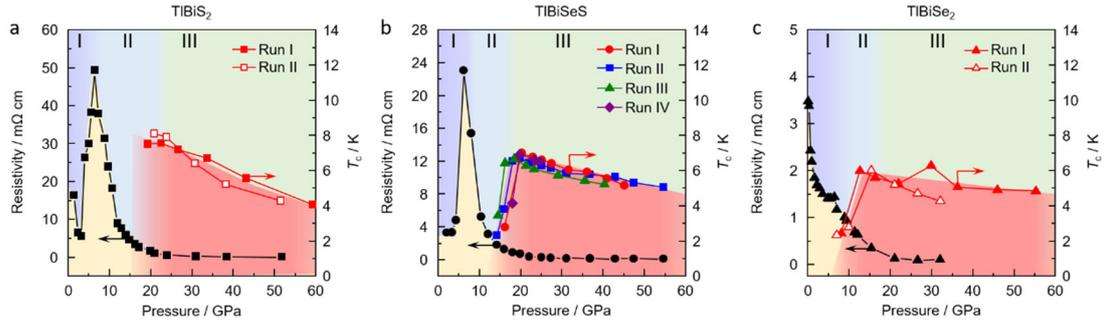

**Figure 4.** Phase diagram for TlBiS$_2$, TlBiSeS, and TlBiSe$_2$. The superconducting $T_c$ and resistivity as a function of pressure at 300 K for TlBiS$_2$ (a), TlBiSeS (b), and TlBiSe$_2$ (c) in different runs. The values of $T_c$ onset were determined from the high-pressure resistivity. The purple, blue, and green region stands for phase I ($R\bar{3}m$), phase II ($C2/m$), and phase III ($P4mm$), respectively.

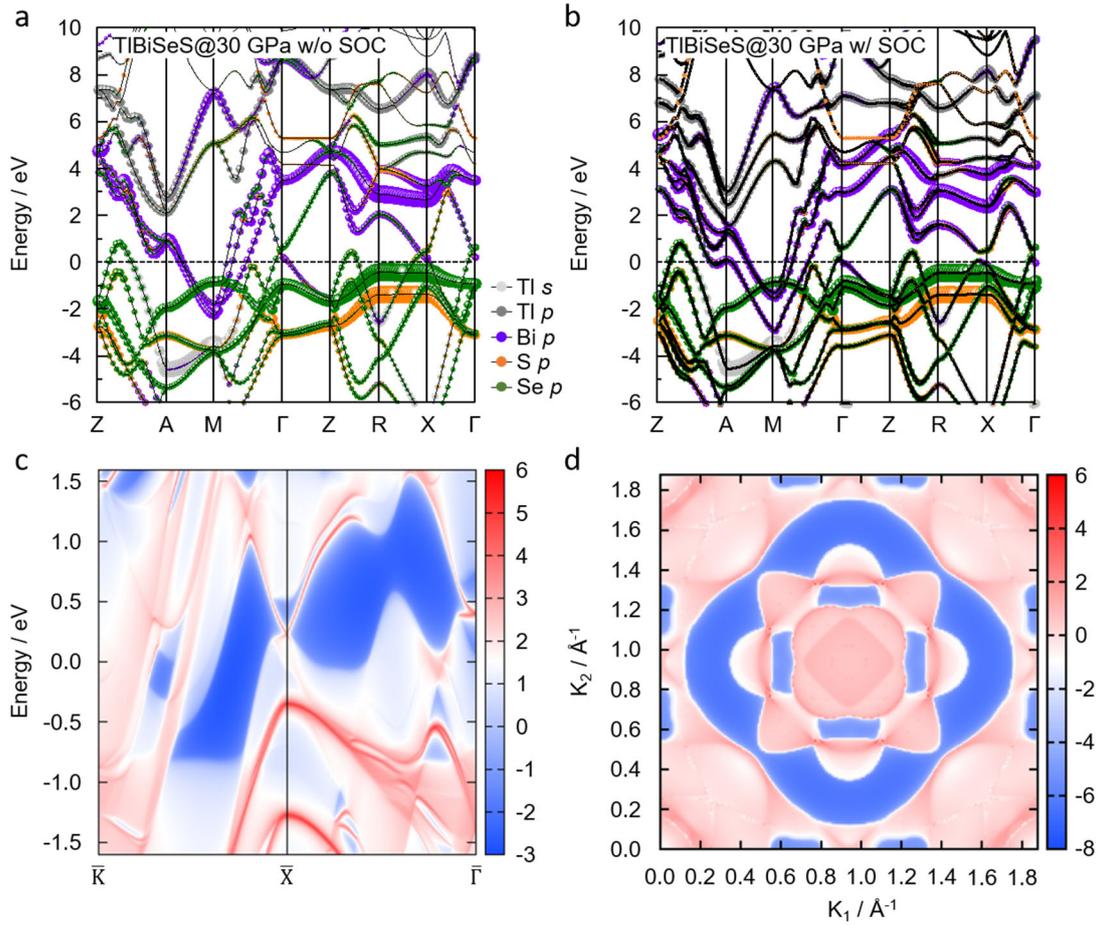

**Figure 5.** Calculated Fermi surface and band structure under high pressure. Orbital-resolved band dispersions near Fermi level for TlBiSeS at 30 GPa (a) without and (b) with SOC calculated by using PBE; (c) Calculated topological edge states of TlBiSeS with SOC; (d)The (111)-surface Fermi surface of TlBiSeS at E = 0 eV.

Supporting Information

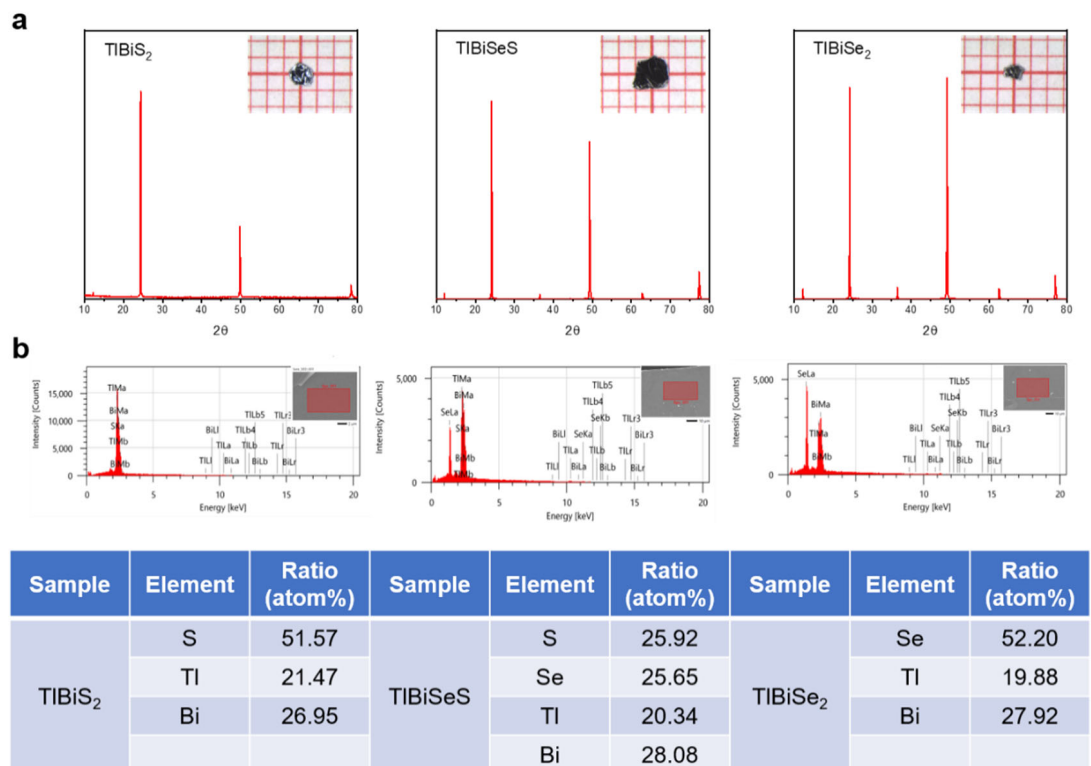

| Sample | Element | Ratio (atom%) | Sample | Element | Ratio (atom%) | Sample | Element | Ratio (atom%) |
|---|---|---|---|---|---|---|---|---|
| TlBiS$_2$ | S | 51.57 | TlBiSeS | S | 25.92 | TlBiSe$_2$ | Se | 52.20 |
| | Tl | 21.47 | | Se | 25.65 | | Tl | 19.88 |
| | Bi | 26.95 | | Tl | 20.34 | | Bi | 27.92 |
| | | | | Bi | 28.08 | | | |

Figure S1. (a) The room temperature powder XRD peaks of TlBiS$_2$, TlBiSeS and TlBiSe$_2$ crystal. Insert pictures are the images of typical TlBiS$_2$, TlBiSeS and TlBiSe$_2$ single crystal synthesized in this work, respectively; (b) The stoichiometry of TlBiS$_2$, TlBiSeS and TlBiSe$_2$ crystal measured by the EDS spectrum. Insert pictures show the crystal used for EDS measurements, respectively. The average compositions derived from a typical EDS measured at several points on the crystal.

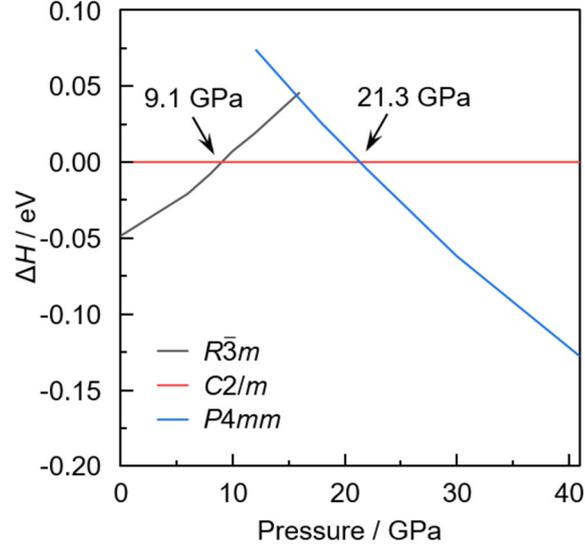

Figure S2. Calculated enthalpies versus pressure of various TlBiSeS structures relative to that of the *C*2/*m* phase.

We have examined structures of TlBiSeS by using our developed structure search method[10; 50; 51]. Structure searches are carried out at 0, 12, 16 and 30 GPa with simulation cells ranging from one to four formula units. Three stable phases were found in the pressure range 0-40 GPa. At 0 GPa, our structure searches simulation successfully reproduced the $R\bar{3}m$ phase, in agreement with previous studies and with the same symmetry as TlBiS$_2$ and TlBiSe$_2$. Upon compression, *C*2/*m* is found to become more stable than the $R\bar{3}m$ phase at 9.1 GPa. At higher pressure, the *P*4*mm* phase is stable above 21.3 GPa.

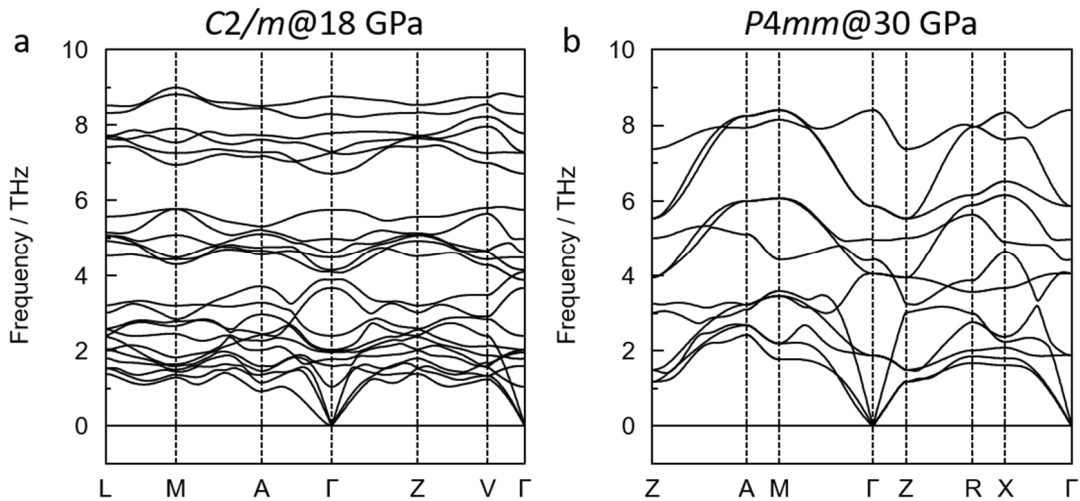

Figure S3. Phonon dispersion of the *C*2/*m*-structure TlBiSeS at 18 GPa (a) and *P*4*mm*-structure at 30 GPa (b), respectively. No imaginary frequency was found for these two structures, indicating dynamical stability of two high pressure phases of TlBiSeS.

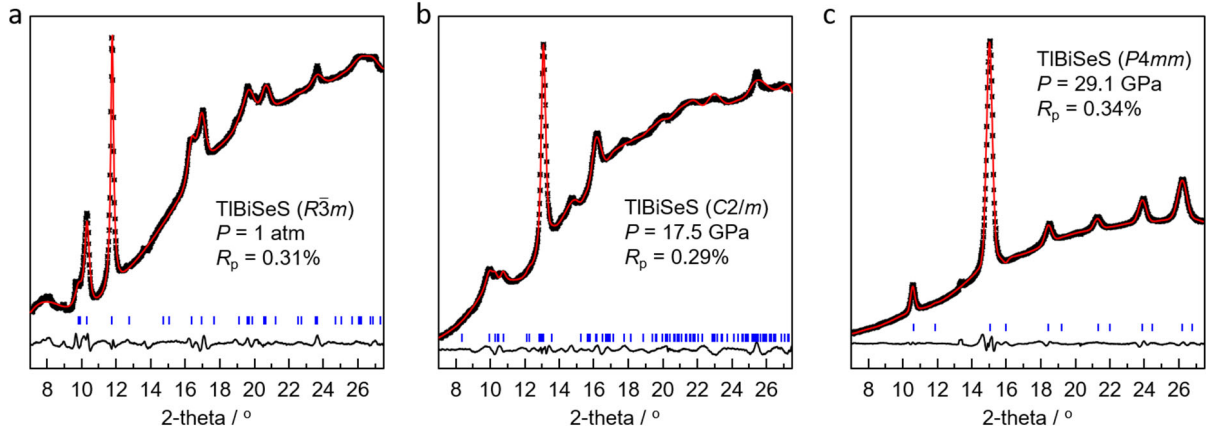

Figure S4. Typical Rietveld refinement results of TlBiSeS under 1 atm (a), 17.5 GPa (b) and 29.1 GPa (c), respectively. The experimental and simulated data were symbolled with black stars and red line. The solid lines shown at the bottom of the figures are the residual intensities. The vertical bars indicate peak positions of the Bragg reflections for TlBiSeS in $R\bar{3}m$(a), $C2/m$ (b) and $P4mm$ (c) space groups.

Table S1. Structural parameters of the three high pressure phases of TlBiSeS at room temperature.

|  | phase I | phase II | phase III |
|---|---|---|---|
| Pressure | 1 atm | 17.5 GPa | 29.1 GPa |
| Crystal system | rhombohedral | monoclinic | tetragonal |
| Space group | $R\bar{3}m$ (166) | $C2/m$ (12) | $P4mm$ (99) |
| a | 4.205(2) | 13.656(1) | 3.350(4) |
| b | 4.205(2) | 3.783(1) | 3.350(4) |
| c | 21.734(1) | 10.938(4) | 6.698(5) |
| α | 90 | 90 | 90 |
| β | 90 | 141.4(1) | 90 |
| γ | 120 | 90 | 90 |
| atoms position | Wyckoff (x y z) | Wyckoff (x y z) | Wyckoff (x y z) |
| Tl | 3a (0,0,0) | 4i (0.1180,0.5,0.5076) | 1a (0,0,0.0916) |
| Bi | 3b (0,0,0.5) | 4i (0.1333,0,0.0157) | 1a (0,0,0.6063) |
| Se | 6c (0,0,0.2383) | 4i (0.8874,0.5,0.7611) | 1b (0.5,0.5,0.8315) |
| S | 6c (0,0,0.2625) | 4i (0.3628,0.5,0.2207) | 1b (0.5,0.5,0.3551) |
| Residuals[a] / % | $R_{wp}$: 0.43% | $R_{wp}$: 0.37% | $R_{wp}$: 0.54% |
|  | $R_p$: 0.31% | $R_p$: 0.29% | $R_p$: 0.34% |

[a] $R_{wp}$ and $R_p$ as defined in GSAS[26]

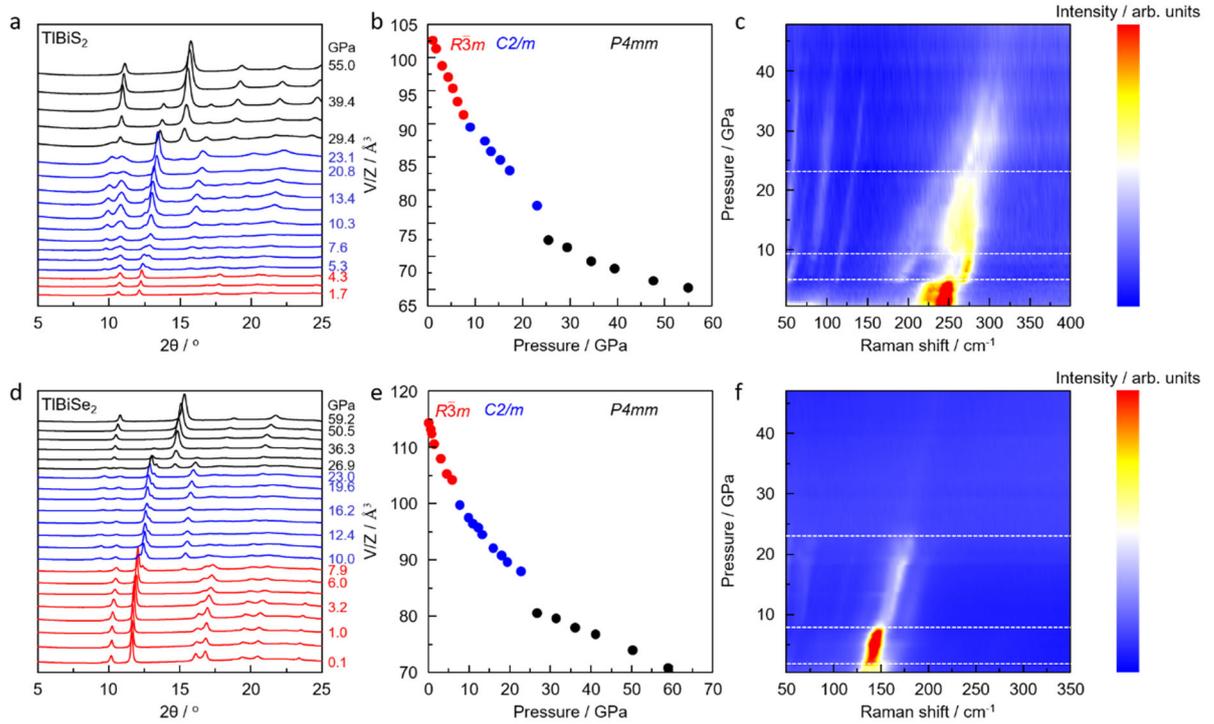

Figure S5. XRD patterns of TlBiS$_2$ (a) and TlBiSe$_2$ (d) under pressure at room temperature with an x-ray wavelength of $\lambda = 0.6199$ Å. The red, blue and black patterns distinguish phase transition from phase I ($R\bar{3}m$) to phase II ($C2/m$) and phase III ($P4mm$), respectively; Pressure-dependence of volume for TlBiS$_2$ (b) and TlBiSe$_2$ (e) in phase I ($R\bar{3}m$), phase II ($C2/m$) and phase III ($P4mm$); Contour color plot of the pressure-dependent Raman spectra of TlBiS$_2$ (c) and TlBiSe$_2$ (f).

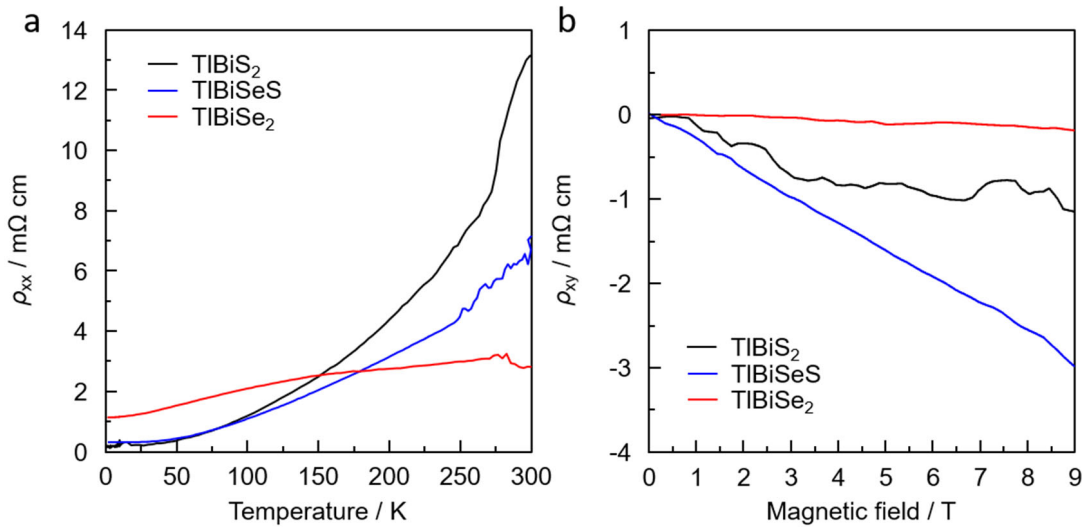

Figure S6. (a) Temperature dependence of the resistivity $\rho_{xx}$ of TlBiS$_2$ (black), TlBiSeS (blue) and TlBiSe$_2$ (red) in the absence of a magnetic field, respectively; (b) the field dependence of $\rho_{xy}$ of TlBiS$_2$ (black), TlBiSeS (blue) and TlBiSe$_2$ (red) at 300 K, respectively.

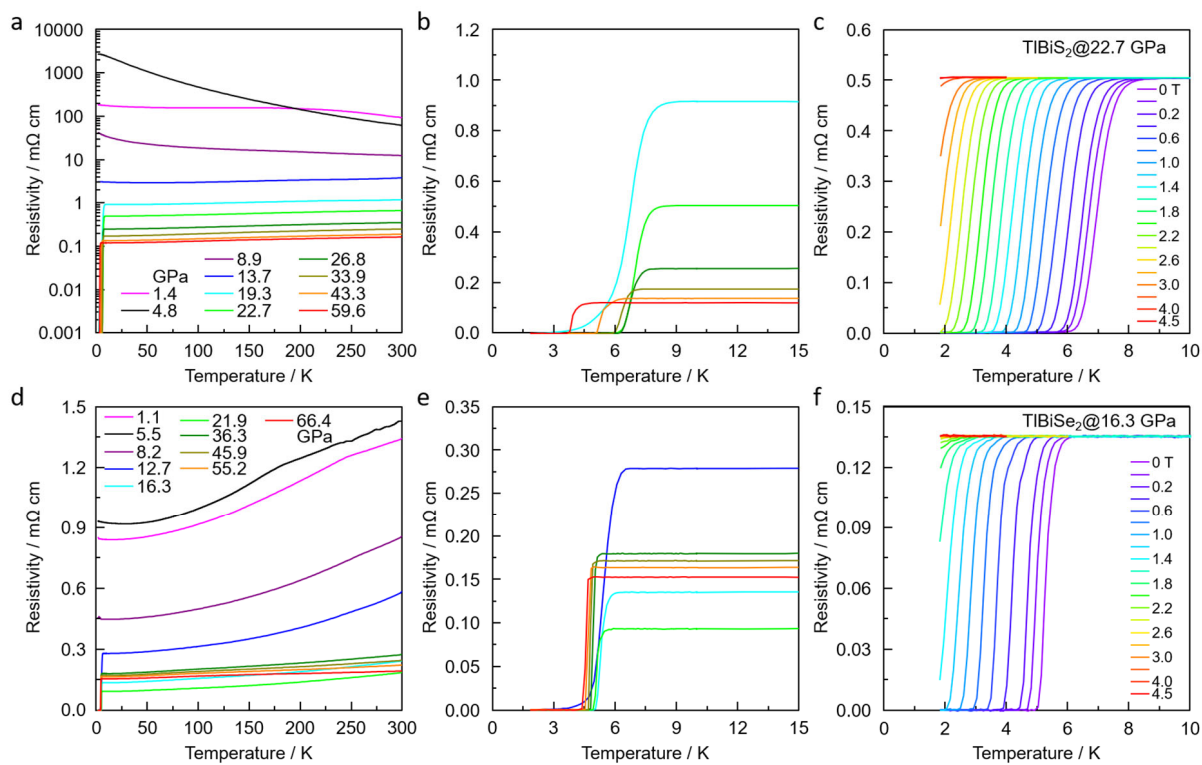

Figure S7. (a) Electrical resistivity of TlBiS$_2$ as a function of temperature for various pressures in run I; (b) Temperature-dependent resistivity of TlBiS$_2$ in the vicinity of the superconducting transition; (c) Temperature dependence of resistivity under different magnetic fields for TlBiS$_2$ at 22.7 GPa; (d) Electrical resistivity of TlBiSe$_2$ as a function of temperature for various pressures in run I; (e) Temperature-dependent resistivity of TlBiSe$_2$ in the vicinity of the superconducting transition; (f) Temperature dependence of resistivity under different magnetic fields for TlBiSe$_2$ at 16.3 GPa.

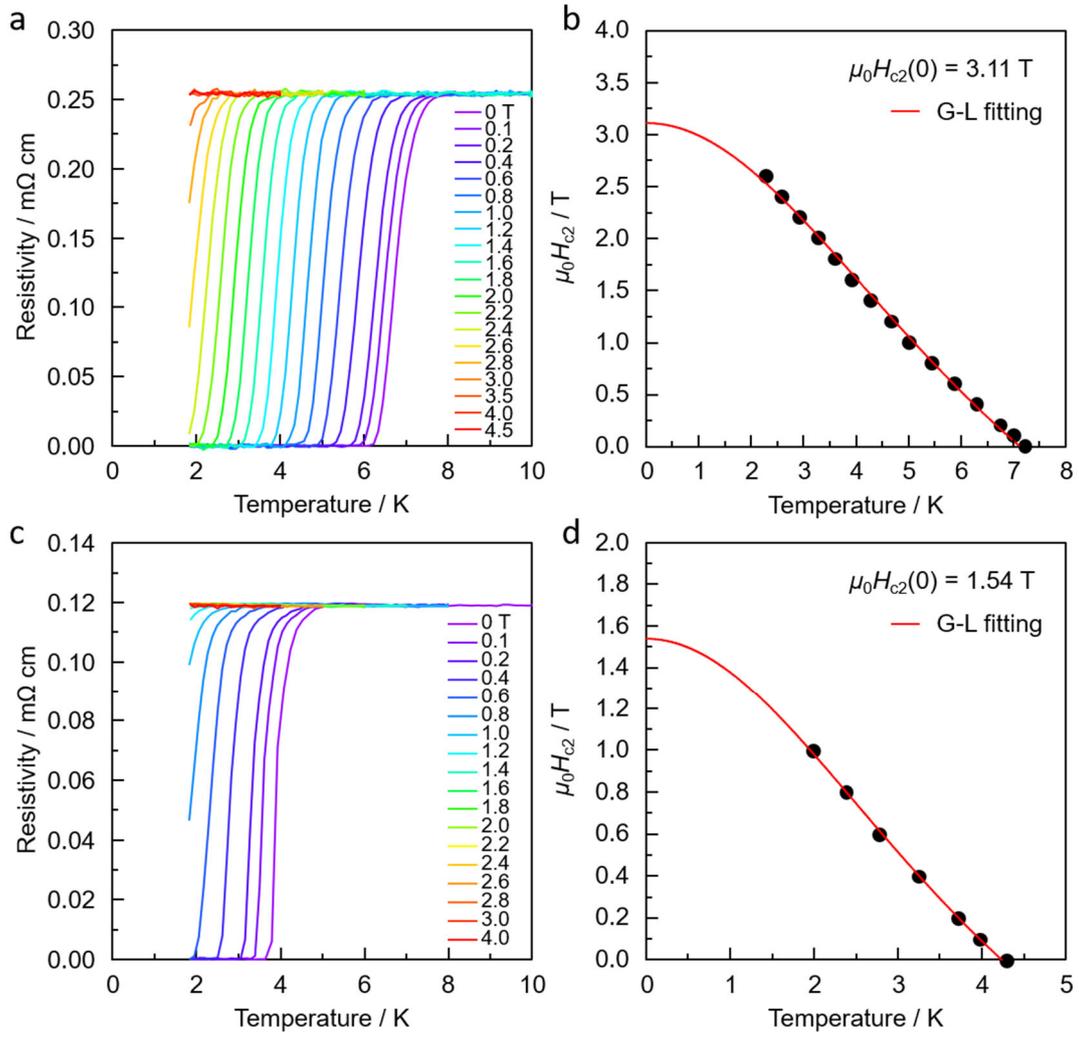

Figure S8. Temperature dependence of resistivity under different magnetic fields for TlBiS$_2$ at 26.8 GPa (a) and 59.6 GPa (c) in run I; Temperature dependence of upper critical field for TlBiS$_2$ at 26.8 GPa (a) and 59.6 GPa (c) in run I. Here, $T_c$ is determined as the 90% drop of the normal state resistivity. The solid lines represent the fits based on the Ginzburg–Landau (G-L) formula.

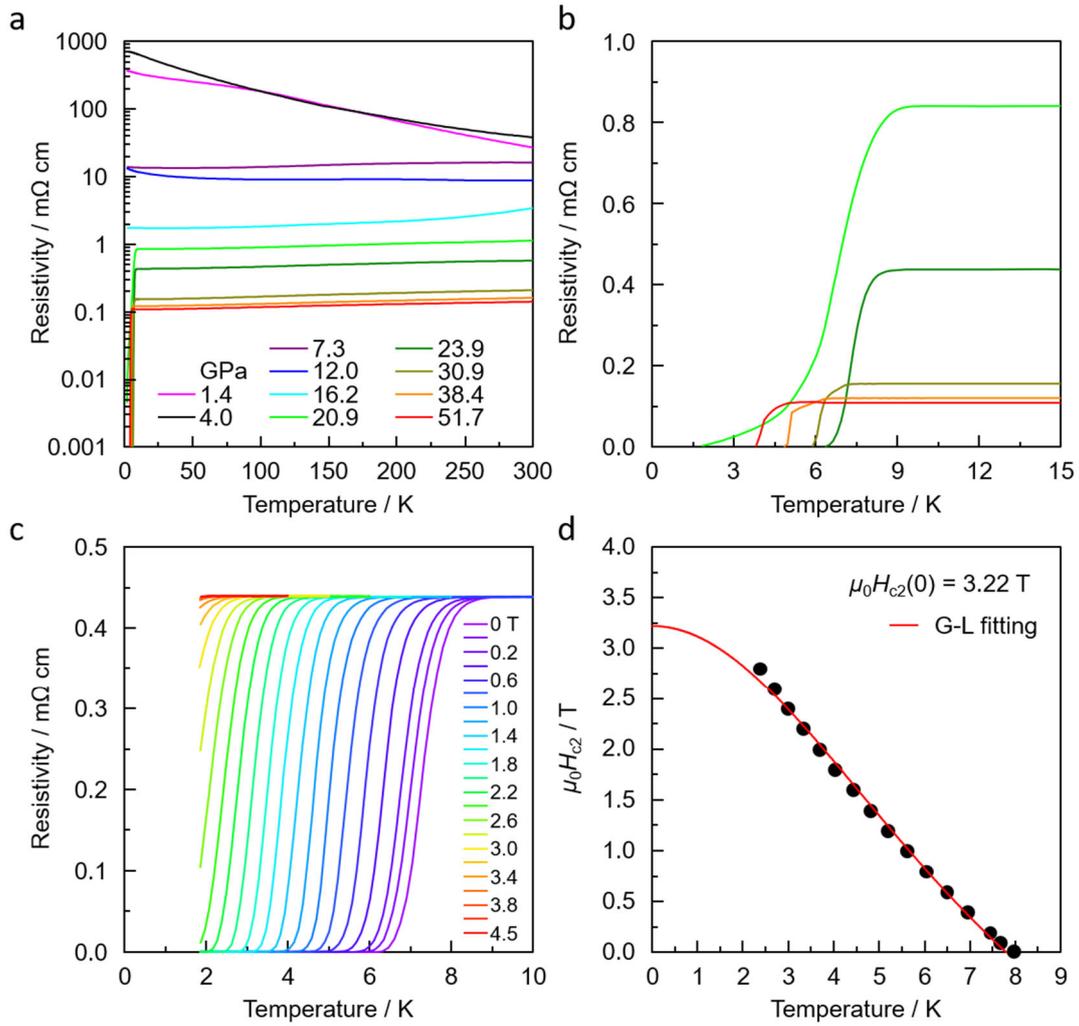

Figure S9. (a) Electrical resistivity of TlBiS$_2$ as a function of temperature for various pressures in run II; (b) Temperature-dependent resistivity of TlBiS$_2$ in the vicinity of the superconducting transition in run II; (c) Temperature dependence of resistivity under different magnetic fields for TlBiS$_2$ at 23.9 GPa; (d) Temperature dependence of upper critical field for TlBiS$_2$ at 23.9 GPa. Here, $T_c$ is determined as the 90% drop of the normal state resistivity. The solid lines represent the fits based on the Ginzburg–Landau (G-L) formula.

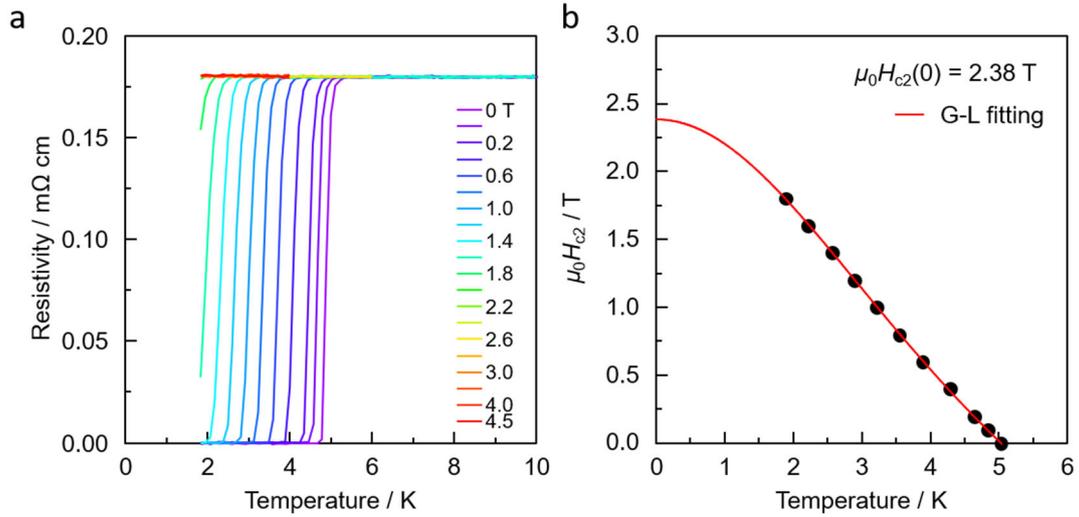

Figure S10. (a) Temperature dependence of resistivity under different magnetic fields for TlBiSe$_2$ at 36.3 GPa in run I; (b) Temperature dependence of upper critical field for TlBiSe$_2$ at 36.3 GPa in run I. Here, $T_c$ is determined as the 90% drop of the normal state resistivity. The solid lines represent the fits based on the Ginzburg–Landau (G-L) formula.

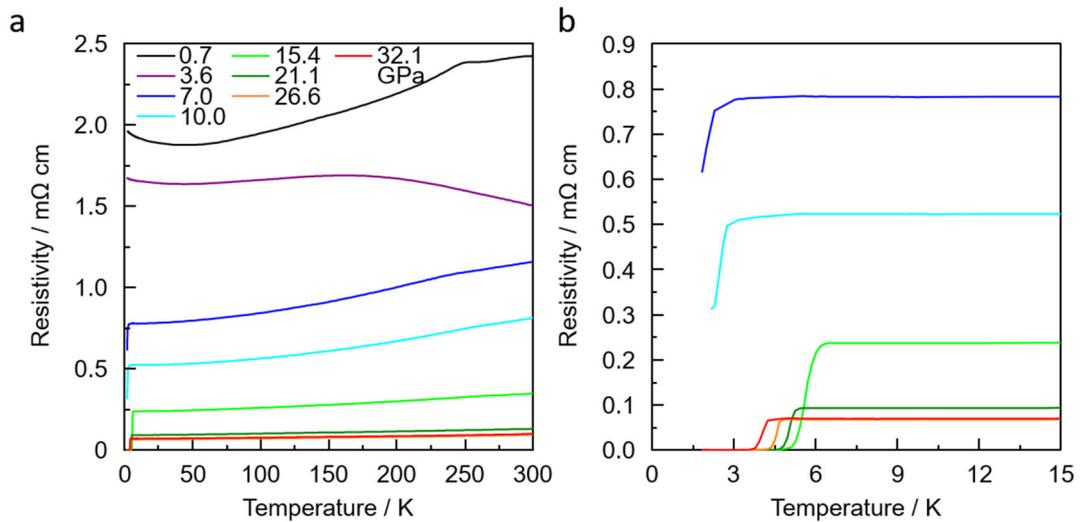

Figure S11. (a) Electrical resistivity of TlBiSe$_2$ as a function of temperature for various pressures in run II; (b) Temperature-dependent resistivity of TlBiSe$_2$ in the vicinity of the superconducting transition in run II.

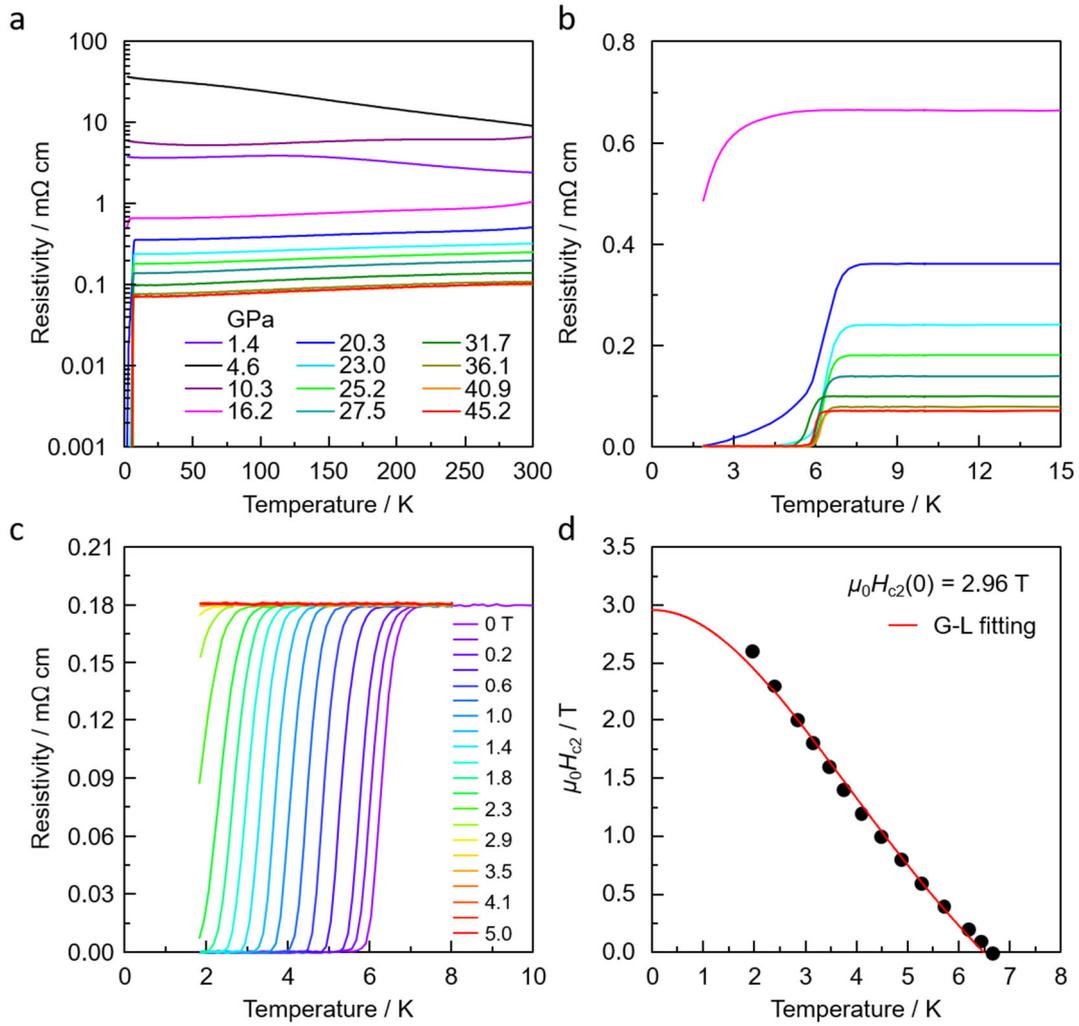

Figure S12. (a) Electrical resistivity of TlBiSeS as a function of temperature for various pressures in run I; (b) Temperature-dependent resistivity of TlBiSeS in the vicinity of the superconducting transition in run I; (c) Temperature dependence of resistivity under different magnetic fields for TlBiSeS at 25.2 GPa; (d) Temperature dependence of upper critical field for TlBiSeS at 25.2 GPa. Here, $T_c$ is determined as the 90% drop of the normal state resistivity. The solid lines represent the fits based on the Ginzburg–Landau (G-L) formula.

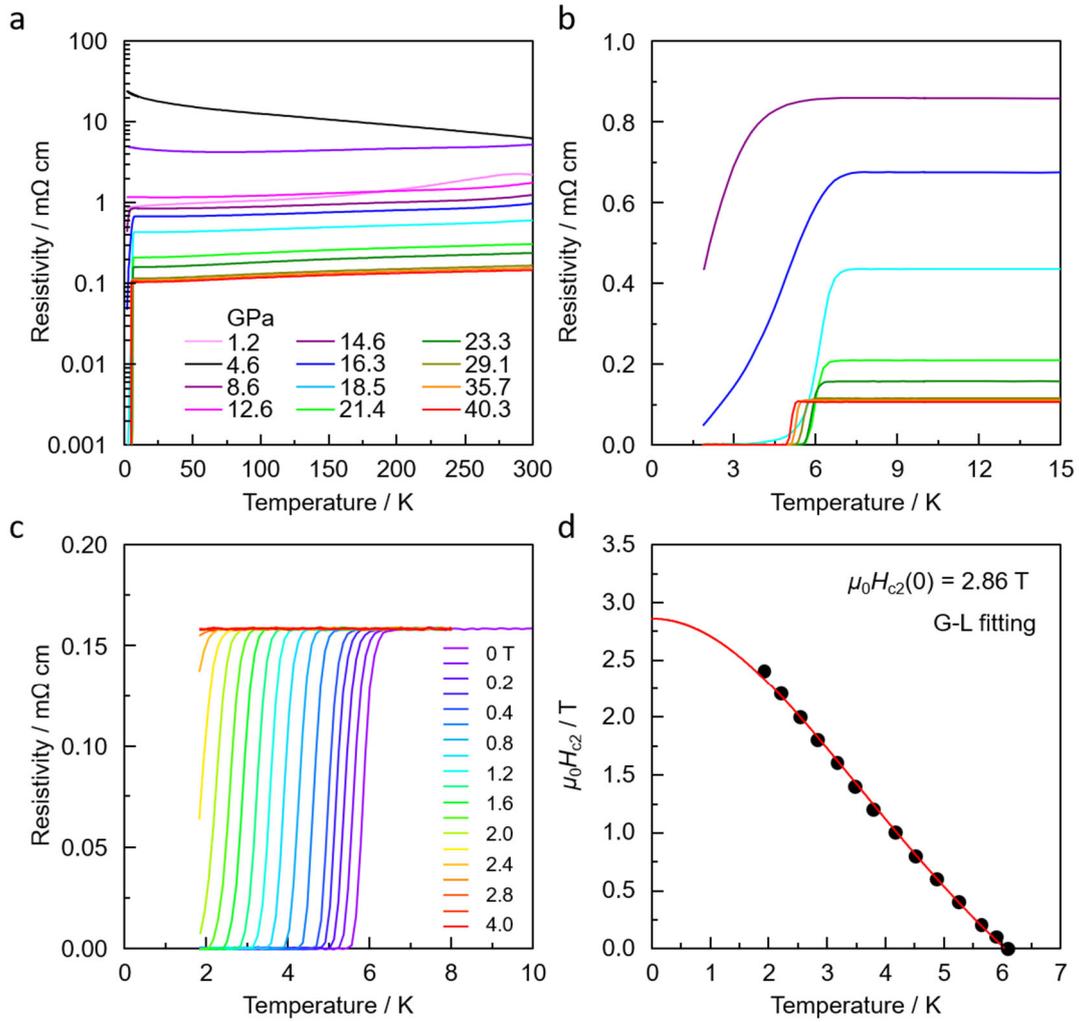

Figure S13. (a) Electrical resistivity of TlBiSeS as a function of temperature for various pressures in run III; (b) Temperature-dependent resistivity of TlBiSeS in the vicinity of the superconducting transition in run III; (c) Temperature dependence of resistivity under different magnetic fields for TlBiSeS at 23.3 GPa; (f) Temperature dependence of upper critical field for TlBiSeS at 23.3 GPa. Here, $T_c$ is determined as the 90% drop of the normal state resistivity.

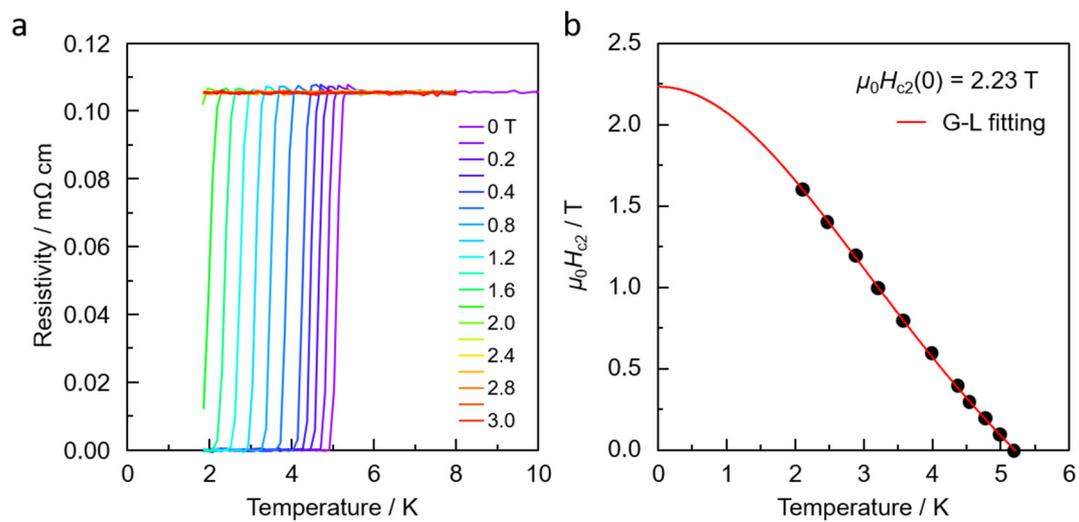

Figure S14. (a) Temperature dependence of resistivity under different magnetic fields for TlBiSeS at 40.3 GPa in run III; (b) Temperature dependence of upper critical field for TlBiSeS at 40.3 GPa in run III. Here, $T_c$ is determined as the 90% drop of the normal state resistivity. The solid lines represent the fits based on the Ginzburg–Landau (G-L) formula.

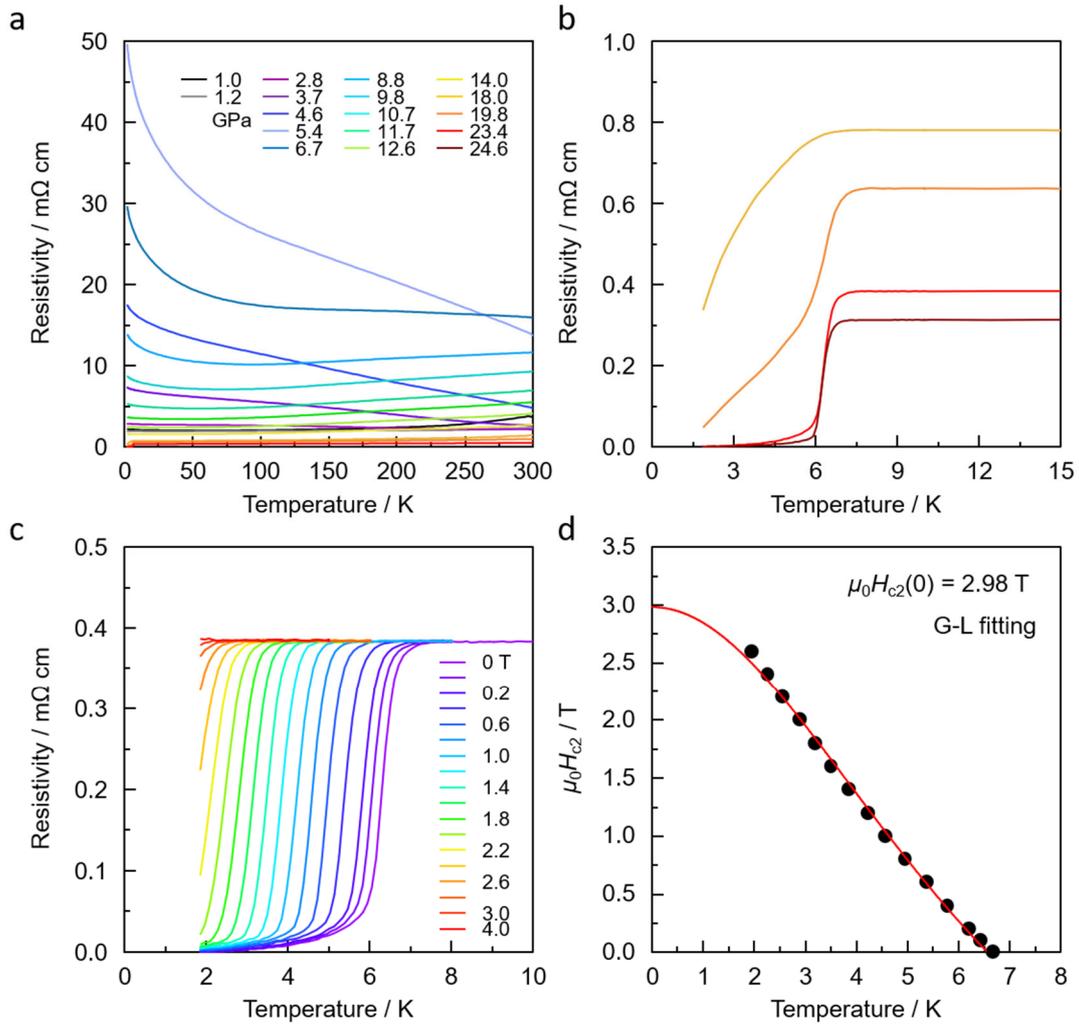

Figure S15. (a) Electrical resistivity of TlBiSeS as a function of temperature for various pressures in run IV; (b) Temperature-dependent resistivity of TlBiSeS in the vicinity of the superconducting transition in run IV; (c) Temperature dependence of resistivity under different magnetic fields for TlBiSeS at 23.4 GPa; (d) Temperature dependence of upper critical field for TlBiSeS at 23.4 GPa. Here, $T_c$ is determined as the 90% drop of the normal state resistivity. The solid lines represent the fits based on the Ginzburg–Landau (G-L) formula.

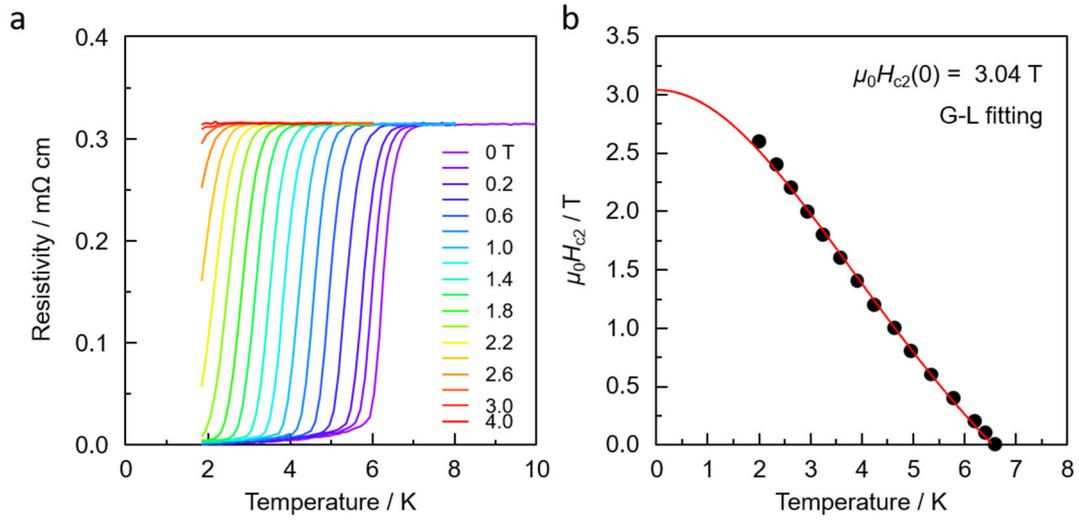

Figure S16. (a) Temperature dependence of resistivity under different magnetic fields for TlBiSeS at 24.64 GPa in run IV; (b) Temperature dependence of upper critical field for TlBiSeS at 24.64 GPa in run IV. Here, $T_c$ is determined as the 90% drop of the normal state resistivity. The solid lines represent the fits based on the Ginzburg–Landau (G-L) formula.

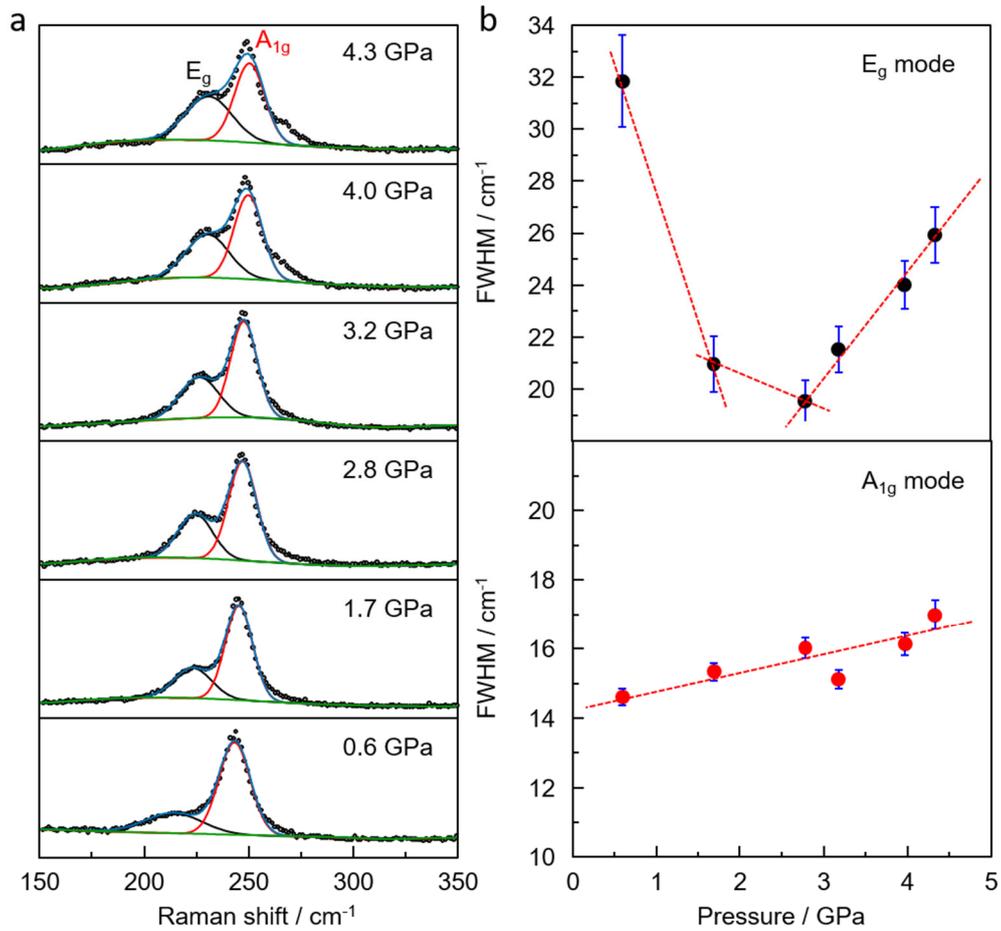

Figure S17. (a) Experimental Raman spectrum of TlBiS$_2$ at various selected pressures; (b) pressure-dependent of FWHM of E$_g$ and A$_{1g}$ mode. The lines (blue, red, black) are the Gaussian fits to the experimental data (black). Green line stands for the baseline with spline interpolation method.

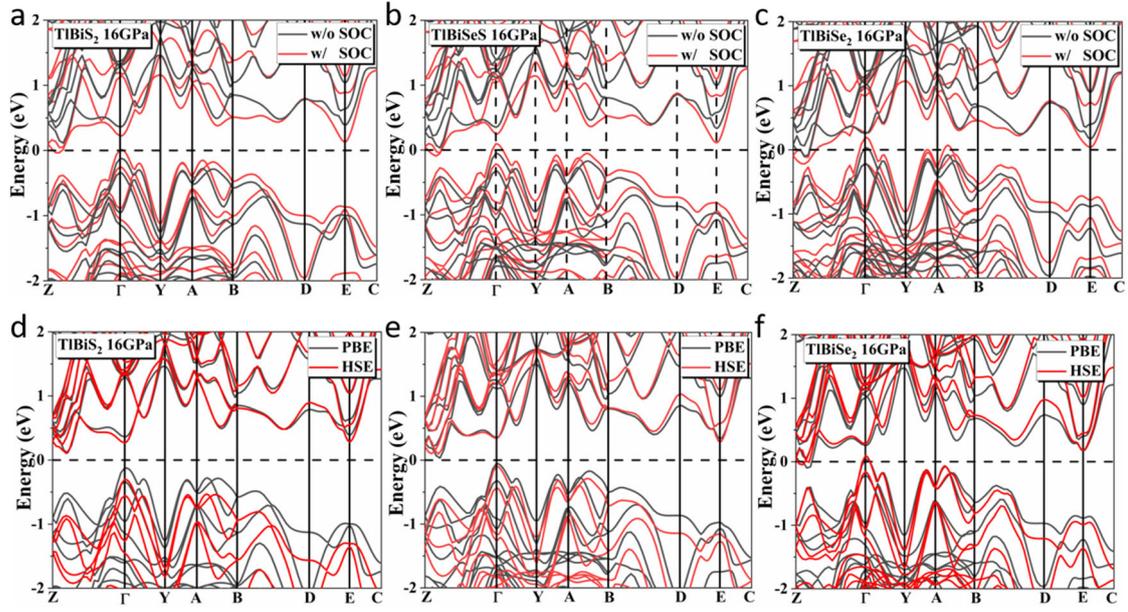

Figure S18. Electronic band structure of TlBS$_2$ (a), TlBiSeS (b) and TlBiSe$_2$ (c) at 16 GPa with/without SOC using the PBE functional; Electronic band structure of TlBS$_2$ (d), TlBiSeS (e) and TlBiSe$_2$ (f) at 16 GPa in absence of SOC using HSE06 Hybrid functional.

**TiBiS₂**

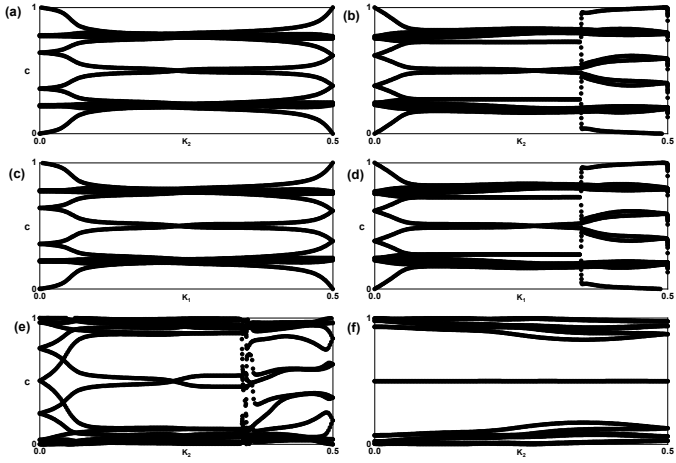

**TiBiSe₂**

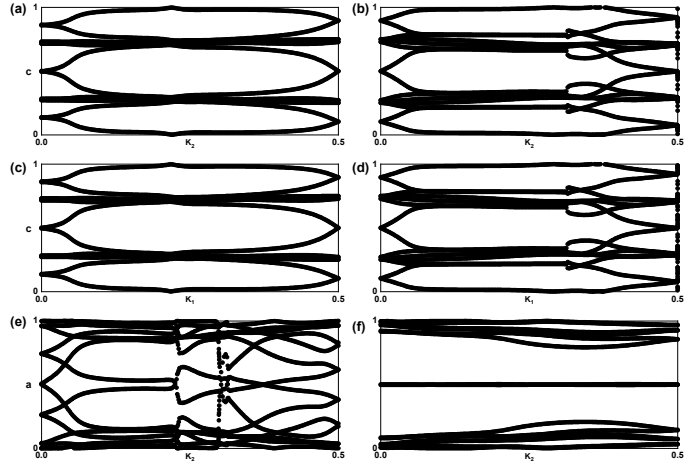

**TiBiSeS**

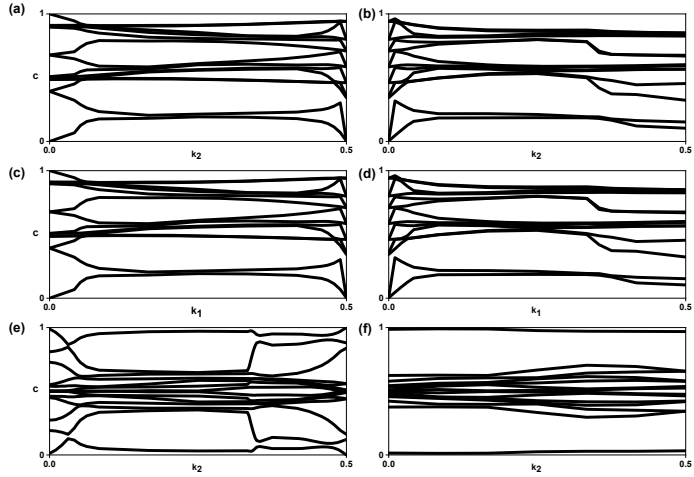

Figure S19. Evolution of Wannier charge centers on six time-reversal invariant momentum planes, (a) k1 = 0.0, (b) k1 = 0.5, (c) k2 = 0.0, (d) k2 = 0.5, (e) k3 = 0.0, and (f) k3 = 0.5 for $TiBiS_2$, $TiBiSe_2$, and $TlBiSeS$ at 30 GPa.

The Z2 topological invariants for these six planes in $TiBiS_2$ at 30 GPa are found to be.
   k1=0.0, k2-k3 plane:          0
   k1=0.5, k2-k3 plane:          0
   k2=0.0, k1-k3 plane:          0
   k2=0.5, k1-k3 plane:          0
   k3=0.0, k1-k2 plane:          1
   k3=0.5, k1-k2 plane:          0

The Z2 topological invariants for these six planes in $TiBiSe_2$ at 30 GPa are found to be.
   k1=0.0, k2-k3 plane:          0
   k1=0.5, k2-k3 plane:          0
   k2=0.0, k1-k3 plane:          0
   k2=0.5, k1-k3 plane:          0
   k3=0.0, k1-k2 plane:          1
   k3=0.5, k1-k2 plane:          0

The Z2 topological invariants for these six planes in TiBiSeS at 30 GPa are found to be.
   k1=0.0, k2-k3 plane:          1
   k1=0.5, k2-k3 plane:          0
   k2=0.0, k1-k3 plane:          1
   k2=0.5, k1-k3 plane:          0
   k3=0.0, k1-k2 plane:          0
   k3=0.5, k1-k2 plane:          0

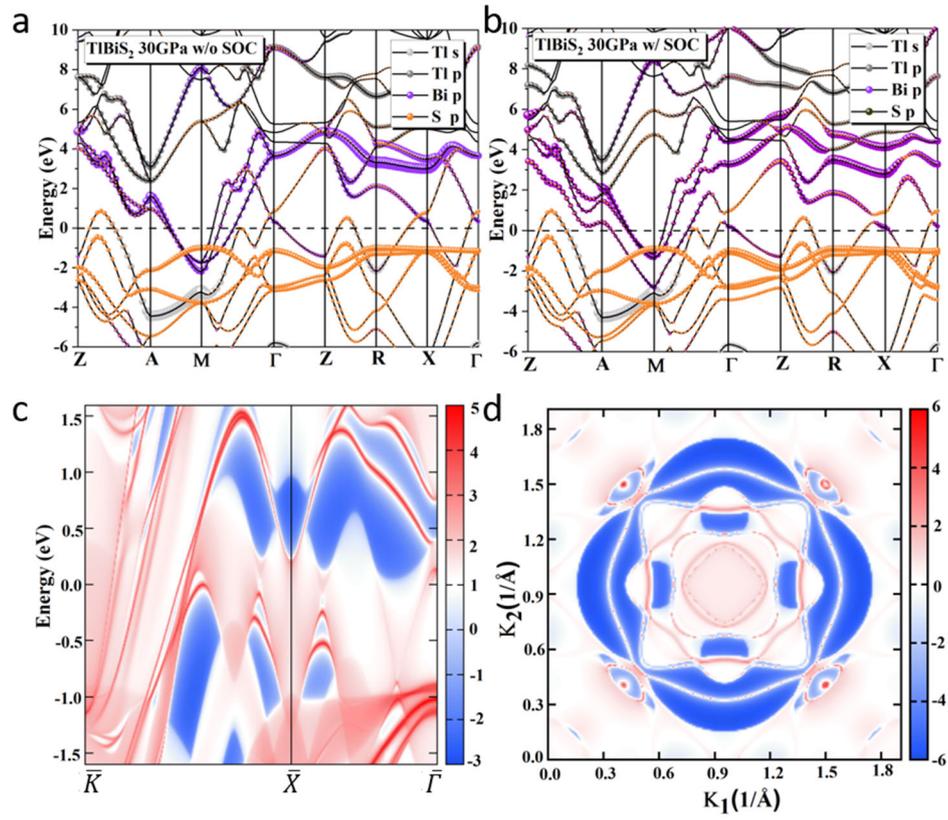

Figure S20. Orbital-resolved band dispersions near Fermi level for TlBiS$_2$ at 30 GPa (a) without and (b) with SOC calculated by using PBE; (c) Calculated topological edge states of TlBiS$_2$ with SOC; (d)The (111)-surface Fermi surface of TlBiS$_2$ at E = 0 eV.

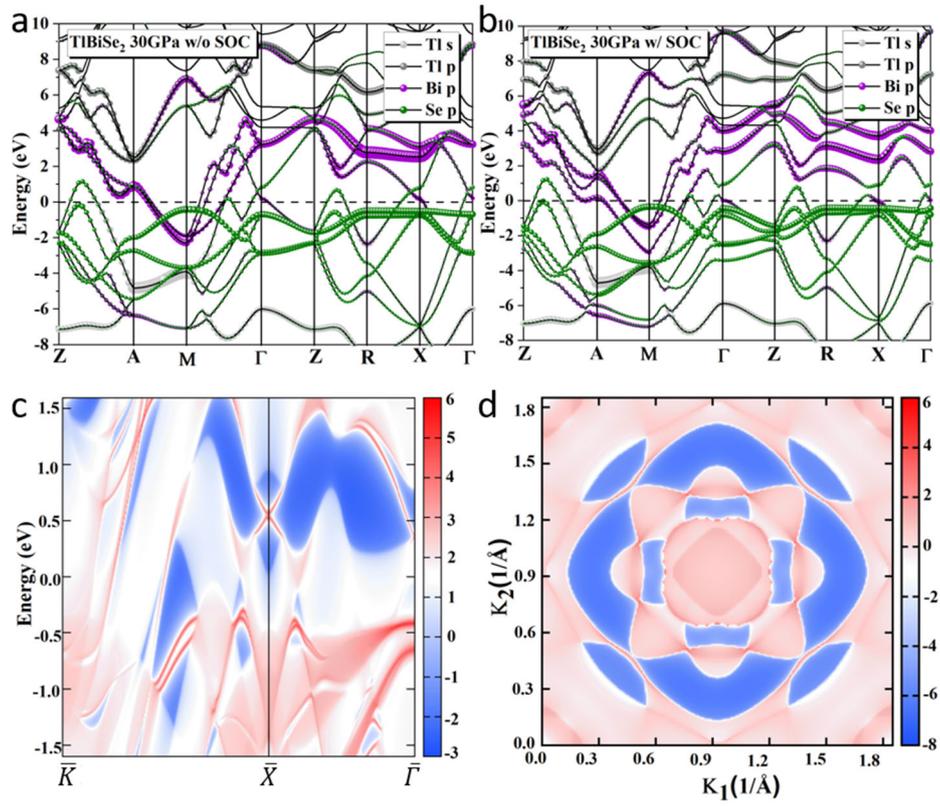

Figure S21. Orbital-resolved band dispersions near Fermi level for TlBiSe$_2$ at 30 GPa (a) without and (b) with SOC calculated by using PBE; (c) Calculated topological edge states of TlBiSe$_2$ with SOC; (d)The (111)-surface Fermi surface of TlBiSe$_2$ at E = 0 eV.

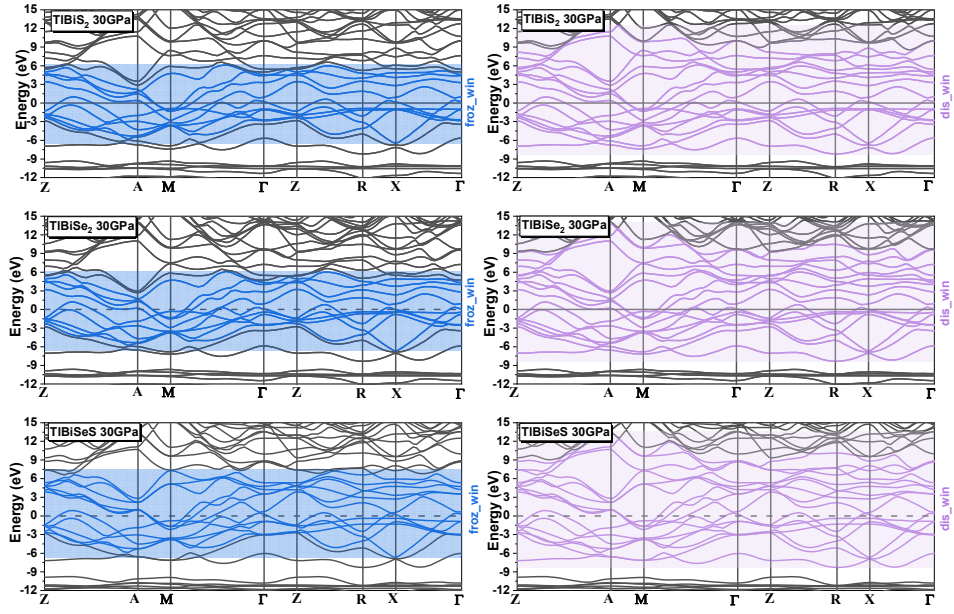

Figure S22. The frozen window froz_win (left) and disentanglement window dis_win (right) of TiBiS$_2$, TiBiSe$_2$ and TiBiSSe at 30 GPa.

Table S2. The parity of the band at X point for three materials TiBiS$_2$, TiBiSe$_2$, and TiBiSSe at 30 GPa. Here, we show the parities of twenty occupied bands, and the lowest unoccupied band. The product of the parities for the twenty occupied bands is given in brackets on the right of each row.

|  | 1 | 2 | 3 | 4 | 5 | 6 | 7 | 8 | 9 | 10 | 11 | 12 | 13 | 14 | 15 | 16 | 17 | 18 | 19 | 20 | 21 |  |
|---|---|---|---|---|---|---|---|---|---|---|---|---|---|---|---|---|---|---|---|---|---|---|
| TiBiS$_2$ | − | − | − | − | − | + | − | − | − | − | − | − | − | − | + | + | − | − | + | +; | − | (−) |
| TiBiSe$_2$ | − | − | − | − | − | + | − | − | − | − | − | − | − | − | + | + | − | − | + | +; | − | (−) |
| TiBiSeS | + | − | + | − | + | − | − | + | + | + | − | + | − | + | − | + | + | − | + | −; | + | (−) |